\documentclass[onecolumn,prd,superscriptaddress,showpacs,floatfix,%
preprintnumbers,nofootinbib,showkeys]{revtex4}  

\usepackage{amsmath,amsfonts}
\usepackage{graphicx}
\usepackage{epsfig,epsf}
\usepackage{hyperref}

\def\beq{\begin{equation}}
\def\eeq{\end{equation}}
\def\bea{\begin{eqnarray}}
\def\eea{\end{eqnarray}}
\def\barr{\begin{array}}
\def\earr{\end{array}}

\def\gev{\, {\rm GeV}}
\newcommand{\e}{\epsilon}

\renewcommand{\c}{\text{c}}
\newcommand{\s}{\text{s}}

\def\gtap{\raisebox{-.4ex}{\rlap{$\sim$}} \raisebox{.4ex}{$>$}}

\usepackage[dvips]{color}
\usepackage[normalem]{ulem}

\definecolor{darkgreen}{cmyk}{1,0,1,0.4}
\definecolor{pink}{cmyk}{0.4,1,0.3,0}

\begin{document}

\title{
Anomalous Higgs Couplings as a Window to New Physics
}

\author{Debajyoti Choudhury}
\email{debajyoti.choudhury@gmail.com}
\affiliation{Department of Physics and Astrophysics, University of Delhi, 
Delhi 110007, India.}

\author{Rashidul Islam}
\email{islam.rashid@gmail.com}
\affiliation{Department of Physics and Astrophysics, University of Delhi, 
Delhi 110007, India.}

\author{Anirban Kundu}
\email{anirban.kundu.cu@gmail.com}
\affiliation{Department of Physics, University of Calcutta,\\
92, Acharya Prafulla Chandra Road, Kolkata 700009, India. }

\date{\today}

\begin{abstract} 

The initial data on the production and decay of the Higgs boson reported significant 
deviations from the Standard Model (SM) expectations, prompting much speculation about 
its couplings to the other particles. Although the latest data has veered towards 
conformity with the SM, there is yet room for a sizable deviation from the SM values 
of the coupling of the Higgs boson with $t\bar{t}$, and to a smaller extent, of that 
with $W^+W^-$ and $ZZ$. Keeping the fluid nature of the data in mind, this opens up 
an interesting avenue to explore regarding unitarity of gauge boson scattering and 
the stability of the electroweak vacuum in the presence of anomalous couplings. We show
that, for some typical benchmark points, unitarity in gauge boson scattering breaks down 
between 1 and 10 TeV. We also show that if there are no new light degrees of freedom, 
the Higgs quartic coupling becomes negative at around the same point, making the 
electroweak vacuum unstable.  Thus, some new ultraviolet completing new physics is 
demanded at that scale to cancel both these anomalous behaviours if such deviations 
from the SM couplings are indeed established.

\end{abstract}



\maketitle

\section{Introduction}

\label{sec:intro}
The recent discovery of a resonance by both the
ATLAS~\cite{ATLASICHEP2012} and the CMS~\cite{CMSICHEP2012}
collaborations at the Large Hadron Collider (LHC) has led to intense
activity. This has been accentuated by the fact that initially both
the groups reported excesses---over the Standard Model (SM)
backgrounds---in multiple channels and concentrated at nearly the same
(125--126 GeV) reconstructed mass. Supported by evidence from the
Tevatron~\cite{Aaltonen:2012qt}, this has naturally led to euphoria in
the community. However, even though the resonance is obviously a
boson, its identification with the long-awaited Higgs particle of the
standard electroweak theory is not yet certain.

While the observed diphoton decay mode excludes the possibility of the
said resonance being a spin-1 particle, other non-trivial assignments
are, as yet, possible. This is notwithstanding the analysis in
Ref.~\cite{cms-cp} where the pure scalar hypothesis is found to be
favoured over the pure pseudoscalar hypothesis for the said resonance.
Furthermore, even if the said assignment is proven, that still does
not uniquely identify the observed particle to be {\em the} SM Higgs
boson, there is still enough room for new physics (NP) resulting
  in modified Higgs couplings with the SM particles.  For one, the
diphoton (the channel permitting the cleanest measurement of the mass)
rate was significantly above the expectations even in late 2012, and
is not yet in full conformity with the SM.  While these are yet early
days to claim a discrepancy, the observed patterns have led to intense
speculations about the nature of this particle and the ramifications
of this discovery for a host of scenarios of physics beyond the
Standard Model (BSM)~\cite{fits-past, fits-recent,
  Banerjee:2012xc}. At this juncture, it must be recognised that the
purported discrepancies, if any, could just be a manifestation
of the inherent uncertainties in QCD calculations, both in the
perturbative and the non-perturbative regimes~\cite{Baglio:2012et}.
On the other hand, ratios of signal rates (such as that between the
diphoton and the four-lepton final states) are relatively free of such
uncertainties and constitute a more robust signature of a deviation
from the SM expectations~\cite{AbdusSalam:2012sy}. A framework for
this has been discussed in Ref.~\cite{1209.0040}, where the scale
factors $g_X$, parametrizing the deviation of Higgs decay width or
production cross section, were defined; the production cross section
$\sigma_{XX}$ for $XX\to h$ or the decay width $\Gamma_{XX}$ for $h\to
XX$ has an extra multiplicative scale factor $g_X^2$ when compared
with the SM predictions.

In view of the state of affairs (compounded by the fact that no other
distinct departure from the SM has been observed at the LHC), the
authors of Ref.~\cite{Banerjee:2012xc} effected an interesting
phenomenological study.  Considering all of the Higgs couplings, whether
tree-level or loop-induced, to be unrelated and free parameters, as
well as allowing for an invisible decay mode for the Higgs, they used
the observations (both low-energy, such as precision electroweak
observables, as well as the recent data) to obtain a best fit to the
same.  In as much as no underlying physics assumptions (other than
Lorentz invariance) were made for this sector ({\em i.e.}, no patterns
were imposed on the anomalous couplings of the Higgs, whether to the
gauge bosons or to the fermions), this constitutes, perhaps, the most
general investigation to the possible nature of physics just beyond
the SM scale.

One might parametrize the Higgs effective coupling to $t\bar{t}$ and
gauge bosons to be
\beq
{\cal L}^{eff}=
e^{i\delta}g_t\dfrac{\sqrt{2} \, m_t}{v} \, h\bar{t}t
+ g_W \, \dfrac{2m^2_W}{v} \, h W_\mu^+ W^{\mu-}
 + g_Z \, \dfrac{m^2_Z}{v} \, h Z_\mu Z^\mu
\,,
    \label{biswa_effective}
\eeq
where $v$ is the vacuum expectation value (VEV) of the Higgs. Within the
SM, $g_t,g_W$ and $g_Z$ all equal unity. Allowing these couplings (as
also others, which are not germane to the discussions here) to vary
independently, Ref.~\cite{Banerjee:2012xc} finds that the ``best fits'',
according to data available in early July 2012, are given by
\beq
\barr{crclcrclcrcl}
({\rm F1}): \qquad & g_t &= &-0.6 & \qquad & g_W & = & 1.2 & \qquad & g_Z & = & 1.6
\\
({\rm F2}): \qquad & g_t &= &-1.3 & \qquad & g_W & = & 1.07 & \qquad & g_Z & = & 1.07
\\
({\rm F3}): \qquad & g_t &= &-1.05 \, e^{0.55 \, i} & \qquad & g_W & = & 1.06 & \qquad & g_Z & = & 1.06 \ .
\\
\earr
    \label{biswa_best_fit}
\eeq
The fits F1 and F2 were performed holding $g_t$ to be
real. Similarly, F2 and F3 demanded custodial symmetry.
The constraint on custodial symmetry breaking, given by
the oblique parameter $T$, is so strong that it is natural to impose
$g_W = g_Z$. 
Note that for these points, all the other parameters were held at their 
best fit values respectively and not at the SM values. 

One must, however, be aware of the fact that the rapid influx of data
makes the best fit points vulnerable to change even over a very short
period of time, and analyses based on the latest data present in
Moriond 2013 are now available in the literature
\cite{falkowski,giardino,ellis}.  For example, Ref.~\cite{falkowski}
obtained $g_V (=g_W=g_Z) = 1.04\pm 0.03$ and $g_t =
1.1^{+0.9}_{-3.0}$, assuming the custodial symmetry and using all
Higgs production and decay data along with electroweak precision
observables; whereas not imposing this symmetry leads
to~\cite{giardino} $g_W = 0.91\pm 0.15$ and $g_Z = 1.02\pm 0.13$.
Note that while our benchmark points F1--F3 may no
longer remain the best fit values, they are still within the 95\%
allowed range. One should also bear in mind that the internal 
disagreements between the various data sets cannot yet be wished away 
and the accumulation of further data can swing the pendulum 
either way. Furthermore, the deviations allowed for by even the Moriond data
are quite significant. In particular, the top quark
Yukawa coupling still has a very large uncertainty and can potentially
be negative. In fact, such a negative coupling, and consequently a
constructive interference between the top-mediated and the
$W$-mediated triangle diagrams (in place of a destructive interference
as in the SM), was touted to be a plausible way out from the apparent
excess in ${\rm Br}(h\to\gamma\gamma)$ in July 2012 data.  The effects
of such anomalous top Yukawa coupling already received attention much
before the discovery of the Higgs boson, {\em e.g.}, in the context of
baryogenesis \cite{whisnant1} or unitarity violation in gauge boson
scattering \cite{whisnant2}. The latter will be particularly relevant
for our subsequent discussion.

As the exact nature of the ``best fit'' would again change once more data is
included in the fit, we do not consider the cases of
Eq.~(\ref{biswa_best_fit}), 
or even the later values, to be sacrosanct, but treat them only as
indicative of such fits. It is worth noting that once custodial
symmetry is imposed, the deviations from the SM, viz. $\delta g_{W, Z}
\; (\equiv g_{W/Z} - 1)$ are much smaller than $\delta g_t$ (this also
holds, albeit weakly, for F1). Although still larger than what naive
dimensional analysis would suggest (for a new physics scale $\gtap
500$ GeV), such $\delta g_{W, Z}$ could, presumably, be the result of
quantum corrections (possibly, though, in a theory that is either
strongly coupled or has a non-trivial ultraviolet completion). The
large change of $\delta g_t$ is, however, a more complicated story and
constitutes the bulk of this paper.

One notes that such a change is also indicated in ATLAS and CMS analyses
\cite{atlas-kappa,cms-kappa} based on their data and the formalism 
developed in Ref.~\cite{1209.0040}. Taking $g_t=g_b=g_\tau$
and $g_W = g_Z$ (so that the custodial symmetry is respected),
the ATLAS Collaboration found, within 68\% confidence limit,
\beq
g_t \in [-1.0,-0.7] \cup [0.7,1.3]\,,\ \ 
g_W \in [0.9,1.0] \cup [1.1,1.3]\,.
\eeq
The CMS Collaboration, on the other hand, found the best fit at 
$(g_t,g_V) \approx (-0.7,0.9)$. However, each of the two collaborations 
analysed only their own data set, and also did not consider the 
possibility that the Higgs could decay into any new particles. Thus, 
the fit in \cite{Banerjee:2012xc} encompasses a wider amount of data. 
With this caveat, it is easy to appreciate the relatively 
minor differences in the fits.

While the process of pinning down the various couplings of the Higgs
continues as data pour in, it is also necessary to subject our
observations to theoretical consistency checks. For example, one
important role of the Higgs boson is to ensure partial wave unitarity
in various $2\rightarrow2$ scattering processes.  The role of
longitudinal $WW$ scattering, for example, in unveiling possible new
physics in anomalous gauge self-interaction, even in the absence of
any well-defined resonance, has been discussed in Ref.~\cite{espriu}.
If the couplings of the Higgs turn out to have non-standard values,
then the fine balance required for unitarity is destroyed, and one has
to set a cut-off scale for the theory~\cite{cheung}. In this work, we
derive values of this cut-off scale for various levels of departure of
the Higgs-fermion-antifermion interactions from their standard
values. Side by side, we also examine the implications of such
modified interaction strengths on the issue of vacuum stability
(essentially arising from the radiatively corrected quartic coupling
potentially turning negative). And, based on the above considerations,
we make some remarks on how the existence of additional particles can
restore balance to the whole scenario, if indeed the recently observed
scalar has anomalous coupling strengths.

The rest of the paper is arranged as follows. In Section \ref{sec:theory}, we discuss
some theoretical issues pertaining to the choice of these best-fit
points; in particular, we would like to spend some time on the point
F3, which includes a nontrivial phase in the top Yukawa coupling.
In Section \ref{sec:unitarity}, we discuss
the unitarity of $WW\to t\bar{t}$ and $ZZ\to t\bar{t}$ scattering with
such benchmark points. The evolution
of the scalar quartic coupling is discussed in Section \ref{sec:rge}. While we do not
go into details about models that can produce such effective couplings,
some relevant remarks are made in Section \ref{sec:avenue}. We summarize and
conclude in the Section \ref{sec:conc}. Some calculational details as well as a compendium
of necessary formulae are put in the appendices.

\section{Some Theoretical Issues}
\label{sec:theory}

Let us first make a few comments on the point F3, where a complex
top Yukawa coupling is indicated.  This immediately raises very
pertinent and interesting questions as to the possible sources of such
an anomalous coupling. While mixing effects (whether in the Higgs
sector or in the fermion sector) can and do cause significant
deviations in the coupling, the magnitude of the deviation is never so
large unless the new states are both very light and have complicated
quantum number assignments. Similarly, such a large anomalous coupling
is not expected from loop-corrections (owing to some as-yet-unobserved
states) unless the said sector couples very strongly to the observed
one\footnote{This observation applies equally to fit F1 as well as
  to $g_t$ of F2.}. In particular, the existence of a non-zero
$\delta$ in Eq.~(\ref{biswa_effective}) ostensibly renders the
Hamiltonian to be non-Hermitian.  Although this, at the first glance,
would seem to lead to non-unitary time evolution, it has been shown
that field theories based on a non-unitary Lagrangian~\cite{lee_wick}
could accommodate a unitary $S$-matrix. This, however, requires a
redefinition of the metric in the Hilbert space, and with such a
modified metric, the $S$-matrix is unitary if {\em all stable
  particles} have a positive-definite norm. Such a redefinition,
however, requires that propagators and Feynman diagrams (indeed, the
entire perturbation theory) be redefined
adequately~\cite{lee_wick_2}. This introduces a whole new panoply of
problems, such as those dealing with electroweak precision
measurements as well as classic tests of quantum electrodynamics such
as the Lamb shift or anomalous magnetic moment of the electron (and,
its cousin, the muon).  This seems to be too big a price to pay and we
shall, hence, turn our attention to simpler alternatives.

As is well-known, such an absorptive part can arise from
loop-corrections within a Hermitian theory if there exists an
intermediate state that can be on-shell. However, the existence of
such a state begs many questions. For one, such particles would
necessarily be light and should have manifested themselves not only in
Higgs decays, but also in other collider processes. This is
particularly so, for, by definition, such a state would be part of an
$SU(2)_L$ doublet, which, in turn, would immediately call for similar
contributions to the absorptive parts of other effective vertices
(with or without the Higgs). Not only this, such a light state should
have been produced directly too. In particular, the $SU(2)_L$
antecedents would have required that they be produced at a clean
environment such as LEP-II (as also the Tevatron). No signs of either
such production, or the inducing of absorptive parts in other
couplings have yet been observed.  Similarly, one cannot ascribe 
this phase due to say, an exchange of an unknown (set of) particle in 
the ``$t$-channel'' of a loop with the on-shell particles being 
some light SM state such as the $b$. Although a phase can 
appear in such a case, it actually encapsulates the final state 
rescattering of the said light particles (i.e. $h \to b \bar b$, 
with $b \bar b$ rescattering, in howsoever complicated a fashion, to 
the final state of interest, viz. $\gamma\gamma$), and has little to do 
with an effective $h t \bar t$ vertex.

Furthermore, the very act of calculating loops with such an ansatz
for the origin of the phase (as attempted in the literature)
is fraught with danger. An effective theory can be
obtained (starting from an ultraviolet completion) only on
integrating out fields more massive than the scale at which the
effective theory is being utilized. By its very definition, then,
the light fields that ostensibly led to the phase cannot be
integrated out and must be included in all loop corrections, whether
for the Higgs production and/or decay, or for processes involving
other particles, such as the $Z$. Such inclusion will, naturally,
lead not only to significant changes in such observables, but, most
often, tend to cancel the effect of the phase (seeing that it is
absent in the complete theory and was but an artefact of a
perturbative calculation).

It should also be realized that, for $m_h \sim 125 \gev$, the top
quark lines at this vertex cannot be on the positive energy
mass-shell. Thus, the application of Cutkowsky rules is not
straightforward; nor is the identification of $\delta$ with the
discontinuity across a cut arising from a physical region singularity.
In other words, the existence of a non-zero $\delta$ in
Eq.~(\ref{biswa_effective}) cannot be motivated from any simple
physics. Indeed, such non-Hermiticity is a subtle issue in quantum
field theory~\cite{lee_wick_2}, and even if such a phase were to occur 
due to some hidden sector exchanges, we cannot include it in an
effective Lagrangian (which is a must for any loop calculations) in 
any straightforward manner.  
We shall, henceforth, consider the
top Yukawa coupling to be real, albeit admitting the possibility of an
anomalous component to it.  Thus, our benchmark point F3 will be
parametrized by $(g_t,\delta) = (-1.05,0)$. Note that this does not
invalidate Ref.~\cite{Banerjee:2012xc}, for when they hold $\delta =
0$, they still find that the best fit requires $g_t \neq 1$ with the
deviation from the SM being substantial\footnote{Indeed, we find the
  admission of a non-zero $\delta$ to be rather unwarranted, given
  that the $\chi^2$-distribution is very flat for $0 < \delta < 1$
  (see Fig.~4 of Ref.~\cite{Banerjee:2012xc}).}.

It is interesting to note that a nonzero phase had been
introduced earlier in the top Yukawa coupling, albeit
in a different context~\cite{whisnant1}. Wishing to
incorporate CP violation in this interaction (motivated
by a desire to address baryogenesis), the authors of Ref.~\cite{whisnant1}
augmented the SM Lagrangian by an effective operator
of the form
\beq
 \delta {\cal L} =   
           c_\phi \,
                           e^{i \xi} \, \bar Q_L \, t_R \, \Phi + h.c
\eeq
where $c_\phi$ denotes a (real) effective coupling owing
its origin to higher-dimension terms. In the unitary gauge,
this yields
\beq
     c_\phi \, \bar t \, \left[ \cos\xi + i \, \sin \xi \, \gamma_5 \right] t \, h
\eeq
over and above the SM term. Clearly, a non-zero $\xi$ leads to CP
violation. This coupling, though, is markedly different from the
ansatz of Ref.~\cite{Banerjee:2012xc}, as it emanates from an
Hermitian effective Lagrangian unlike in the other case. Furthermore,
the pseudoscalar term (which, essentially, is the only one 
to see a non-zero value of the phase $\xi$) 
in the coupling above contributes only
incoherently to $h \to \gamma \gamma$ and is, thus, of little
consequence (at least within the effective theory paradigm).

An anomalous top Yukawa coupling (even if real) 
brings in its own complications.
Within the SM, all couplings are dictated by gauge
invariance\footnote{For example, the Yukawa couplings are uniquely
given in terms of the masses.}. While deviations are indeed possible
once one enlarges the ambit of the theory, gauge invariance would
require that these either be associated with higher-dimensional
effective operators, or be the consequence of mixings between states
(were new states to be admitted). Each of these eventualities would
imply correlated deviations in other couplings, and, on occasions, the
introduction of new ones. Any uncorrelated deviation, such as that of
Eq.~(\ref{biswa_effective}) can only be the result of an additional
term in the Lagrangian of the form
\beq
  {\cal L}_{\rm eff} = {\cal L} + {\cal L}_{\rm anom} \qquad
  {\cal L}_{\rm anom} = (g_t -1) \, \dfrac{\sqrt{2} \, m_t}{v}h\bar{t}t
   \label{our_effective}
\eeq
where, for simplicity, we have chosen $\delta = 0$. Such a term, of
course, explicitly breaks $SU(2)_L \otimes U(1)_Y$. While its
inclusion may seem to militate against the gauge dogma, note that
Eq.~(\ref{our_effective}) could just represent the relevant part of the
BSM physics, with other terms being hidden for unknown reasons. Thus,
the breaking of gauge invariance might be an artefact of restricting
ourselves to be close to the augmented SM, 
which acts only as a low-energy effective theory,
while gauge invariance is again restored when we go to the full
theory at a high energy. In the effective theory, due to the apparent
loss of gauge invariance, the mass and the Yukawa coupling of the
fermions, in particular the top quark, get decoupled, and this
apparent loss has other profound implications.  As is well known,
unitarity in gauge boson scattering (in particular, the longitudinal
modes) is inextricably linked to gauge invariance. While any loss of
unitarity due to ${\cal L}_{\rm anom}$ could, in principle, be
restored on inclusion of other terms in ${\cal L}_{\rm eff}$, the
scale at which such a loss is seen (if one considers ${\cal L}_{\rm
anom}$ alone) would point to the scale of the new theory that
underlies such a deviation. Similarly, the existence of ${\cal L}_{\rm
anom}$ would have non-trivial consequences for the renormalization
group evolution of the couplings in the theory as well for
considerations such as the stability of the vacuum.


\section{Unitarity Bounds}
\label{sec:unitarity}

\subsection{Unitarity and $g_{W,Z}$}
    \label{sec:unitarity_W}

In a phenomenological study of the Higgs boson, while all its
couplings could be varied independently~\cite{Banerjee:2012xc}, it
makes sense to concentrate on the dominant ones. Within the SM, these
are the ones with the top quark and the weak gauge bosons. Maintaining
the Lorentz structures to be identical to those within the SM, these
can be parametrized as in Eq.~(\ref{biswa_effective}),
but categorically with $\delta = 0$.

Clearly the effective Lagrangian in Eq.~(\ref{biswa_effective}) would
have non-trivial effects on a host of scattering processes, notably on
$V_1 V_2 \to V_3 V_4$ where $V_i = W^\pm, Z$.  As is well-known,
partial wave unitarity for such scattering processes depends crucially
on the couplings being those mandated by $SU(2)_L \otimes U(1)_Y$
invariance alongwith renormalizability. Thus, $g_{W,Z} \neq 1$ could,
in principle destroy the same for, say, $W^+_L W^-_L \to W^+_L
W^-_L$. This particular scattering proceeds through a set of seven
Feynman diagrams, namely a four-point contact interaction, two
$s$-channel diagrams mediated by the $\gamma$ and the $Z$ (or, in the
unbroken symmetry phase, by the $W_3$), two analogous $t$-channel ones
and, finally, one each of $s$-- and $t$--channel Higgs-mediated
diagrams.  With the trilinear (quartic) gauge boson vertices scaling
as $k^1$ ($k^0$) where $k$ is a typical momentum transfer, and the
polarization vector for the longitudinal vector boson going (for large
$k$) as $\epsilon_\mu \sim k_\mu / m_W$, it is obvious that each of
the individual pure-gauge diagram contributions to the amplitude goes
as ${\cal M}_i \sim s^2 / m_W^4$. The gauge theory antecedents of the
vector-boson self-couplings ensure that the leading terms cancel
identically leaving behind a $s/ m_W^2$ behaviour. Once the
Higgs-mediated diagrams are included, even the ${\cal O}(s)$
contributions cancel, and on integrating the remaining terms over the
phase space, one obtains a cross section in consonance with the
Froissart bound\footnote{Although the presence of a massless photon in
the $t$-channel results in a collinear singularity, this does not
violate the Froissart bound. Indeed, this singularity disappears (as
it should) when higher order corrections are taken into
account.}. Clearly, this cancellation is contingent upon the Higgs
couplings being just so, and allowing for $g_W \neq 1$ would result in
additional ${\cal O}(s \, \delta g_W/m_W^2)$ contributions from the
Higgs-mediated diagram to the amplitude resulting in a bad high-energy
behaviour.
\begin{figure}[!h]
\centering
\includegraphics[width=0.45\textwidth]{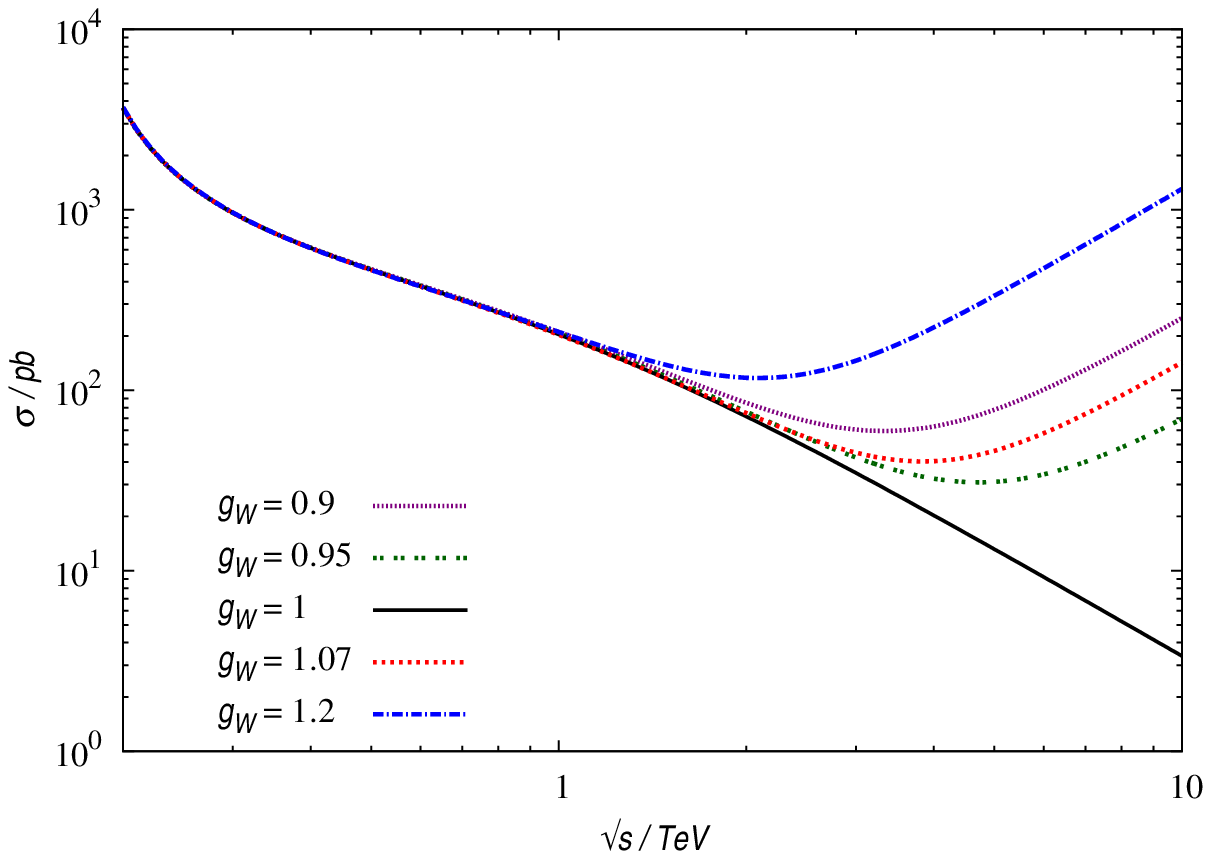}
\includegraphics[width=0.45\textwidth]{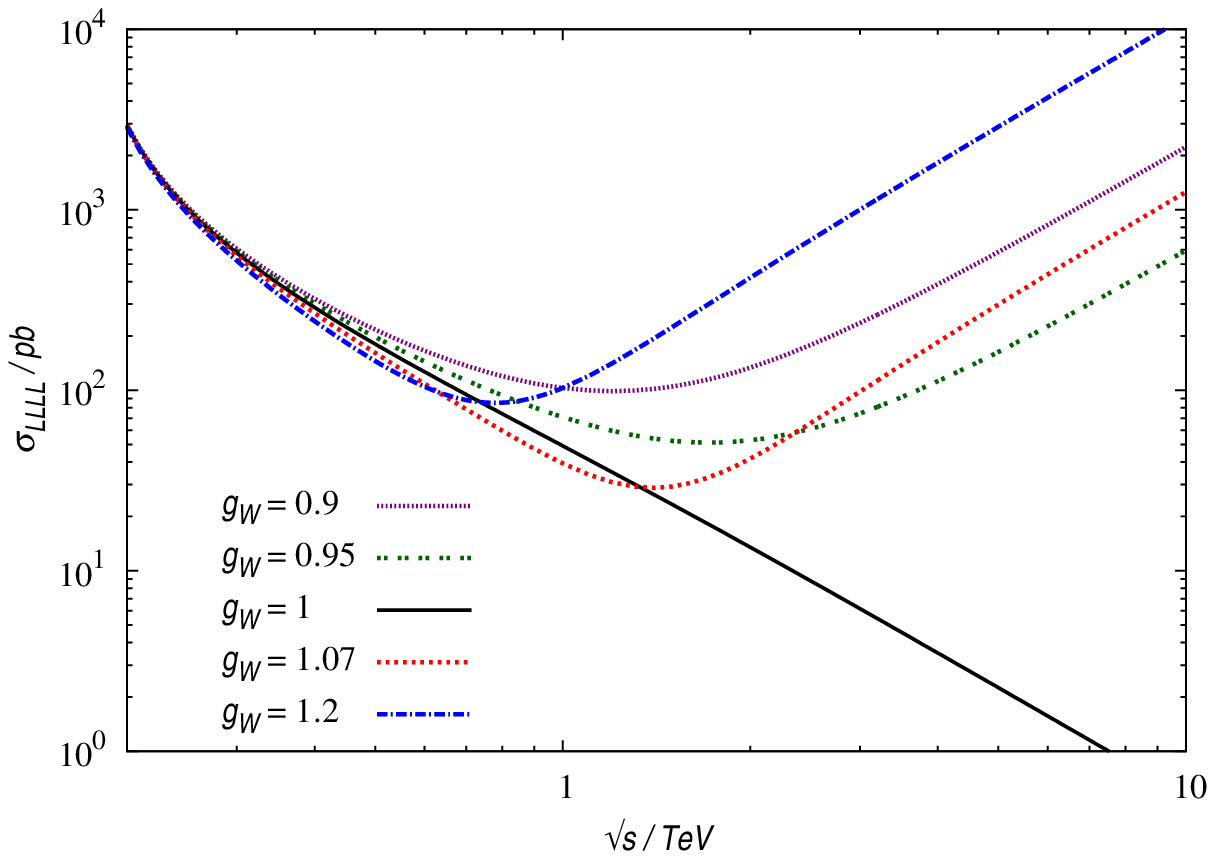}
\caption{\small \em The cross section for $W^+ W^- \to W^+ W^-$
as a function of the CM energy, on imposition of a cut
  $10^\circ \leq \theta \leq 170^\circ$ on the scattering angle.
The left (right) panels refer to unpolarized and $W_L^+ W_L^- \to W_L^+
  W_L^-$ scattering respectively. The
  individual curves refer to different values of the $WWh$ coupling $g_W$
  as normalized to the SM value (see Eq.~\ref{biswa_effective}).}
\label{fig:4W}
\end{figure}
In Fig.~\ref{fig:4W}, we show the consequent behaviour of the cross
sections for a few representative values of $g_W$.  Note that even for
small values of $|g_W-1|$ that are allowed at present, the cross
section grows anomalously  and hence it is easy to
ascertain that such a theory loses unitarity at a few TeVs at
best\footnote{It is instructive to note that the loss of unitarity
  occurs not only for $g_W > 1$ 
  (corresponding to best fit with custodial symmetry \cite{falkowski}), 
but also
  for $g_W < 1$ (best fit without custodial symmetry
    \cite{giardino}).}  and a new theory needs to be around, and, by
implication, {\em within the reach of the LHC}.

It might be argued, though, that such a deviation in $g_W$ could well
be accompanied by others in the gauge boson self couplings, evoking
memories of a non-linearly realized symmetry, or at the very least,
higher-dimensional terms in an electroweak chiral Lagrangian. While it
seems plausible that such correlated deviations could preserve
unitarity, it can be seen that simultaneous restoration in all
possible channels is difficult to achieve within the ambit of
phenomenologically acceptable deviations \cite{forthcoming}. However,
even if this were to be possible, constraints appear from another
sector that we now turn to. This is of particular importance as the
deviations $\delta g_{W,Z}$ in the fits F2 and F3 are relatively
small and could shrink
further once more data is taken into account.

\subsection{Unitarity and $g_{t}$}
    \label{sec:unitarity_top}

As already mentioned, of the SM particles, the Higgs couples with an
unsuppressed strength only to the weak gauge bosons and the top.  We
have already discussed the consequences of deviations to the former
and, now, concentrate on the latter. In analogy to the discussion in the
preceding section, this coupling plays a crucial role in processes such
as $W^+ W^- \to t \bar t$, to which the following diagrams contribute:

\begin{figure*}[!ht]
\centering
\includegraphics[width=\textwidth]{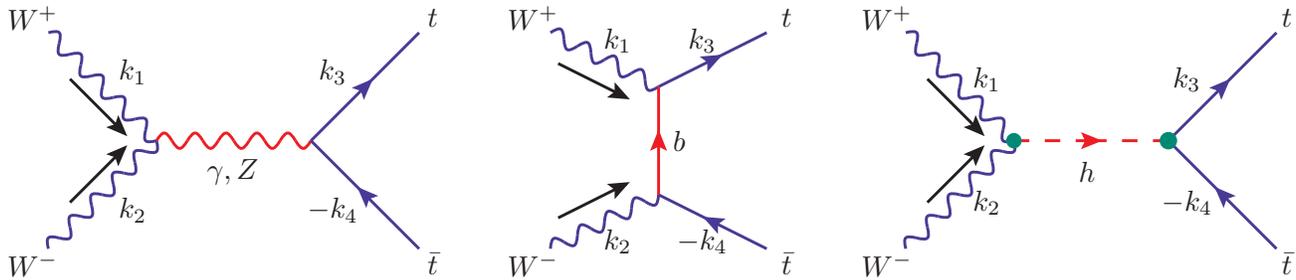}
\caption{\small \em Diagrams contributing to the process $W^+W^- \to t\bar t$.}
\label{fig:WWtt}
\end{figure*}

As can be ascertained from arguments mirroring those in the preceding
section, the amplitude that grows most strongly with energy pertains
to $W_L^+ W_L^-$ annihilation to $t \bar t$. Indeed, the Higgs diagram
contribution goes as ${\cal M}_h \propto g_t \, g_W \, m_t \, \sqrt{s}
/ m_W^2$ for $\sqrt{s} \gg m_t$. If the coupling $g_t$ deviates from
the SM value, the cancellation of the leading term with the non-Higgs
diagrams would be imperfect and the amplitude would grow with energy,
thereby violating the Froissart bound at some scale. While it may be
argued that it is only the combination $g_t \, g_W$ that comes into
play, note that the $\delta g_W$ needed for the fits can neither
compensate for the required $\delta g_t$ nor is such a large deviation
consistent with $WW$ scattering. Similarly, large deviations in the $W
t b$ vertex can be ruled out from the measurements of single-top
production at the Tevatron~\cite{Peters:2012wf} and the
LHC~\cite{LHC_single_top}, as well as from B physics observables such
as the mass difference of neutral B meson eigenstates.

This study is best done in terms of the partial wave amplitudes defined
as
\[
a_\ell \equiv \dfrac{1}{32 \, \pi} \int_{-1}^1 d \cos\theta \;
       P_\ell(\cos\theta) \; {\cal M}(s, \cos\theta; \{m_i, g_i\})
\]
where ${\cal M}$ is the Lorentz invariant amplitude, $\theta$ is the scattering
angle and $P_\ell(x)$ the Legendre functions. Unitarity demands that
\[
  |Re(a_\ell)| < \dfrac{1}{2}\,,~~ \forall \ell
\]
and it is the $l = 0$ amplitude $a_0$ that gives the strongest bound.
In particular, the most sensitive probe is given by the amplitude
for the particular helicity combination
\[
  a_0(0,0,1,1) \equiv a_0(W^+_L W^-_L \to t_+ \bar t_+) \ ,
\]
with the case for $a_0(0,0,-1,-1)$ being identical.
Denoting the velocities of the particles in the
center-of-mass frame by $\beta_W$ and $\beta_t$, one
obtains\footnote{The details of the calculations are given in Appendix A.}
\bea
\label{pwa-WW}
a_0(0,0,1,1) &=& \dfrac{- g^2 \, m_t \, \sqrt{s}}{128 \, \pi \, m_W^2}
\left[ \dfrac{\zeta_W }{\beta_W \, \beta_t} [\beta_W(1-\beta^2_W) - 2a_W\beta_t] \right.
\notag\\
& &\left. \hskip2cm +  \dfrac{\zeta_W }{2 \, \beta_W \, \beta_t \, a_W}
    [\beta_t(1+\beta^2_W) + a_W\beta_W(1-\beta^2_W) - 2a^2_W\beta_t] \,
      \ln\dfrac{a_W - 1}{a_W + 1} \right. \notag\\
& &\left. \hskip2cm + 2 \, g_t \, g_W \, \beta_t \,
            \dfrac{ s - 2 \, m^2_W}{s - m_h^2}
  \right]
    \label{a0_for_WWtt}
\eea where $a_W = (s - 2 m_W^2 - 2 m_t^2)/(\beta_W \beta_t s)$ and
$\zeta_W = |V_{tb}|^2 + |V_{ts}|^2 + |V_{td}|^2 = 1$.  We have assumed
here that the gauge couplings of the top quark are
unaltered\footnote{Note that a significant variation from $\zeta_W =
  1$ is strongly disfavoured by constraints from flavour physics.}
from those in the SM. While no direct measurement of the $Z t \bar t$
vertex is available, once one considers the $W t b $ vertex to be in
consonance with the SM (also indicated to be so by a host of
observables such as single top production, top decays as well as
$B$-meson phenomenology), custodial symmetry mandates that the $Z t
\bar t$ coupling should also be as postulated within the SM. The collinear singularity that
  appears in the large-$\sqrt{s}$ limit---attested to by the
  logarithmic term---is identical to that within the SM and disappears 
  once higher order corrections are taken into account. Any violation
  of unitarity is, then, proportional to  the deviation of 
  the product $g_t \, g_W$ from unity.

\begin{figure*}[!ht]
\centering
\includegraphics[width=\textwidth]{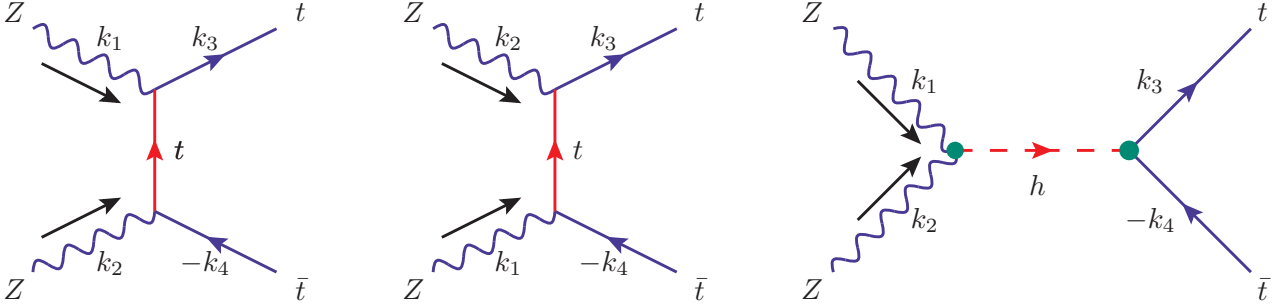}
\caption{\small \em Diagrams contributing to the process $ZZ \to t\bar t$.}
\label{fig:ZZtt}
\end{figure*}

In a similar vein, we can consider $ZZ \to t\bar t$, to which the
diagrams of Fig.~\ref{fig:ZZtt} contribute.
Once again, $a_0(0,0,1,1)$ proves to be the most sensitive probe.
Denoting the coupling of the left--(right--) handed top states with the
$Z$ by $g^{Zt}_L$ ($g^{Zt}_R$), this is given by
\bea\label{pwa-ZZ}
a_0(0,0,1,1) &=&
\dfrac{- m_t \, \sqrt{s}}{32 \, \pi \, m_Z^2}
\left[  \dfrac{(g^{Zt}_L)^2 + (g^{Zt}_R)^2}{\beta_Z \beta_t} [\beta_Z(1-\beta^2_Z) - 2a_Z\beta_t] \right. \notag\\
& &
\left. +
      \dfrac{(g^{Zt}_L)^2 + (g^{Zt}_R)^2}{2 \, \beta_Z \beta_t a_Z} [\beta_t(1+\beta^2_Z) +
a_Z\beta_Z(1-\beta^2_Z) - 2a^2_Z\beta_t] \ln\dfrac{a_Z - 1}{a_Z + 1} \right. \notag\\
& & \left. + \dfrac{4 \,  g^{Zt}_L g^{Zt}_R  }{\beta_t}
+  \dfrac{g^{Zt}_L g^{Zt}_R }{\beta_Z \beta_t a_Z} [\beta_t(1+\beta^2_Z) - 2a_Z\beta_Z]\ln\dfrac{a_Z + 1}{a_Z - 1}\right.
\notag\\
& &
\left. \hskip1cm + g_t g_Z \dfrac{g^2 }{2\c^2_W}
  \dfrac{s - 2 \, m_Z^2}{s - m_h^2} \beta_t   \right]
    \label{a0_for_ZZtt}
\eea
where $a_Z = (s - 2 m_Z^2)/(\beta_Z \beta_t s)$. 

\begin{figure*}[!ht]
\centering
\includegraphics[width=0.45\textwidth]{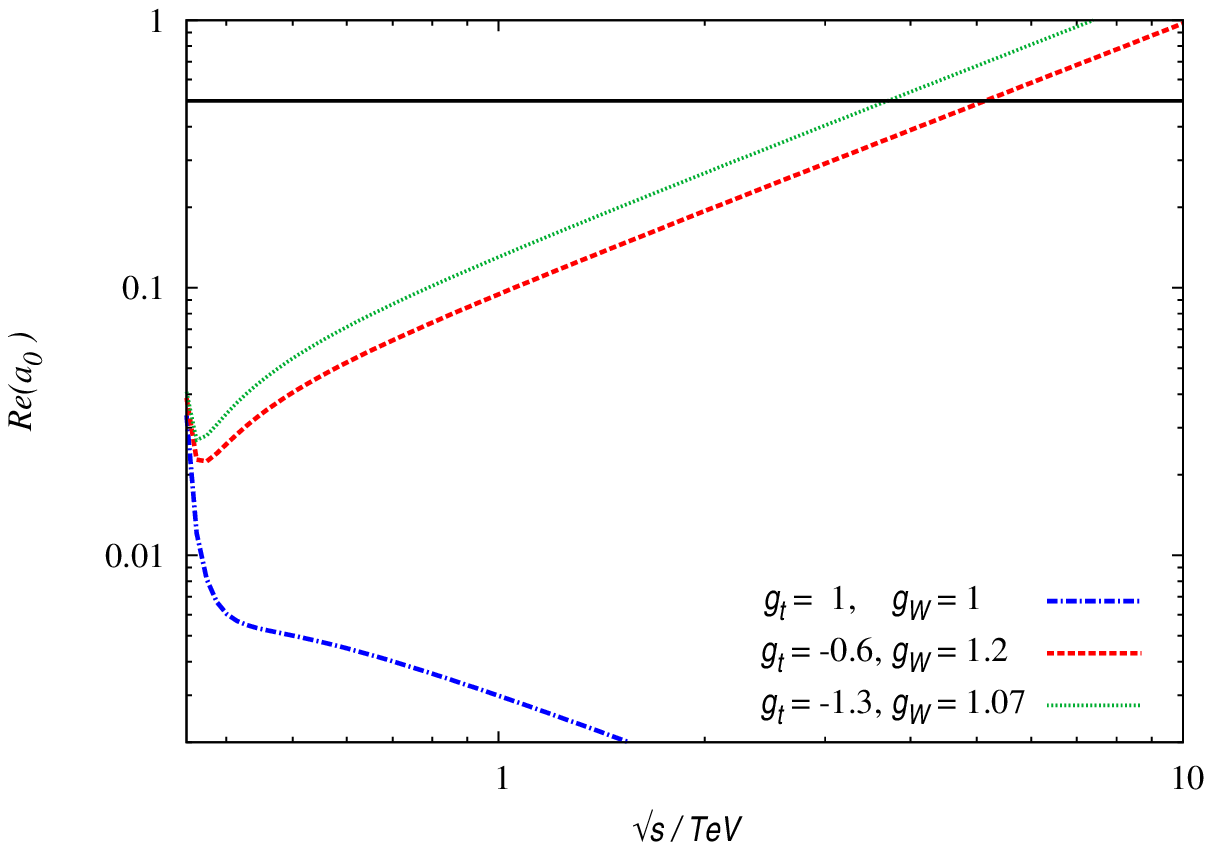}
\includegraphics[width=0.45\textwidth]{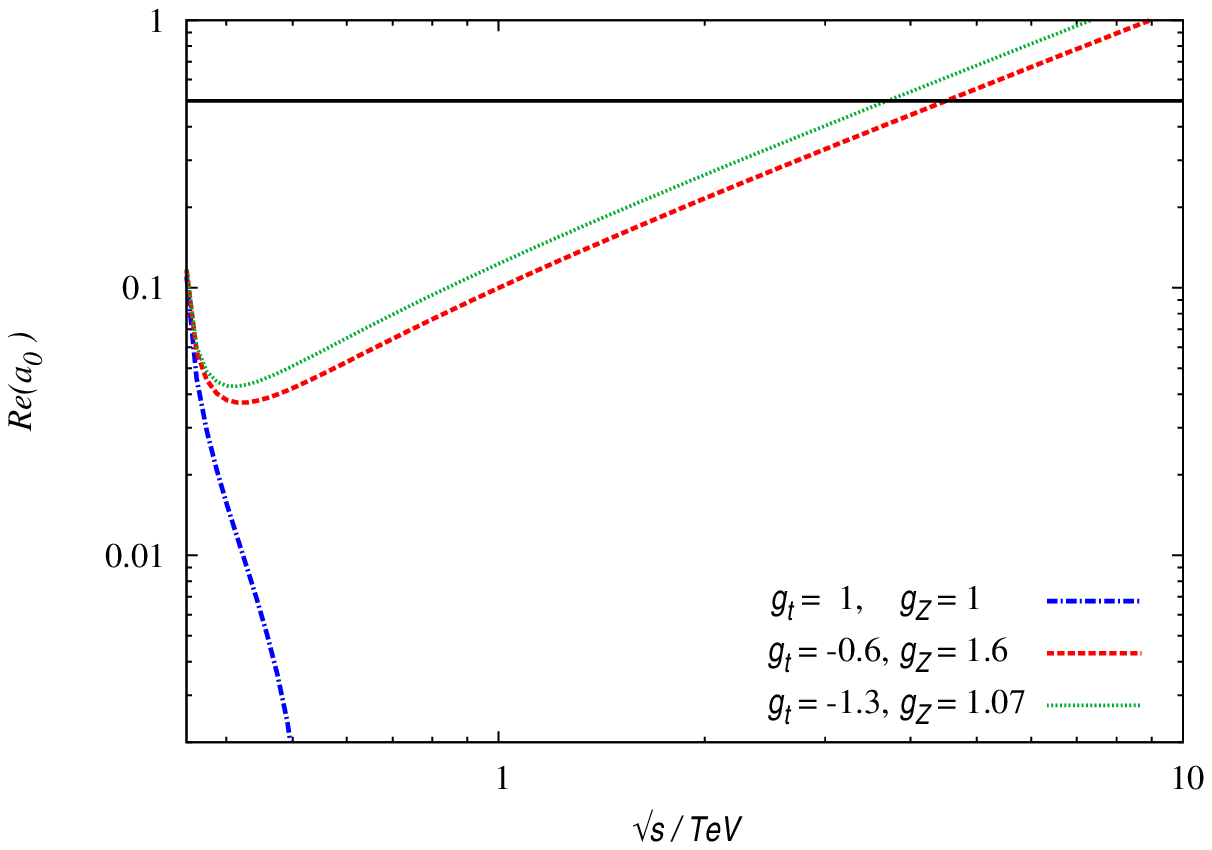}
\caption{\small \em The partial wave amplitude
as a function of the center-of-mass energy. The left (right) panel
corresponds to $W^+_L W^-_L \to t_+ \bar t_+$ ($Z_L Z_L \to t_+ \bar t_+$).
The curve corresponding to $g_t=g_{W/Z}=1$ reflect the SM.
The black solid line denotes the upper limit from unitarity.}
\label{fig:pwave}
\end{figure*}

In Fig.~\ref{fig:pwave}, we show the variation of
the aforementioned $a_0$ with the center-of-mass energy.
As expected, a deviation of the couplings from the SM 
values cause a significant change in the magnitude of $Re(a_0)$. Indeed,
for the most favourable cases of Ref.~\cite{Banerjee:2012xc}
unitarity would be violated at $\sqrt s \gtrsim  4$ TeV, 
while for the  more recent fits \cite{falkowski,giardino},
this would occur at  $\sqrt s \gtrsim  10$ TeV.
In other words,
this indicates the maximal energy scale of the effective theory, beyond
which a new theory must be operative.

It might be argued, though, that much of the unitarity violation
exhibited in Fig.~\ref{fig:pwave} may be caused by the shifts in $g_W$
and $g_Z$. As discussed in the preceding section, such deviations are
strongly disfavoured by considerations involving gauge boson
scattering. Indeed, it can be explicitly checked that the violation of
unitarity owes itself to a negative value for the product $g_t g_W$
(engendered by a negative $g_t$).  Furthermore, the particular values
chosen for the anomalous couplings were dictated by the `best fits'
corresponding to a set of data that might soon be overwhelmed by new
data. In view of this, it is worthwhile to examine the consequences of
having a nonzero $\delta g_t$ alone, while maintaining all other
couplings to their SM values. Indeed, as an examination of 
Eqs.~(\ref{a0_for_WWtt}) and  (\ref{a0_for_ZZtt}) suggests, the extent of 
unitarity violation is determined solely by  $|g_t \, g_{W/Z} - 1|$.

\begin{figure*}[!ht]
\centering
\includegraphics[width=0.45\textwidth]{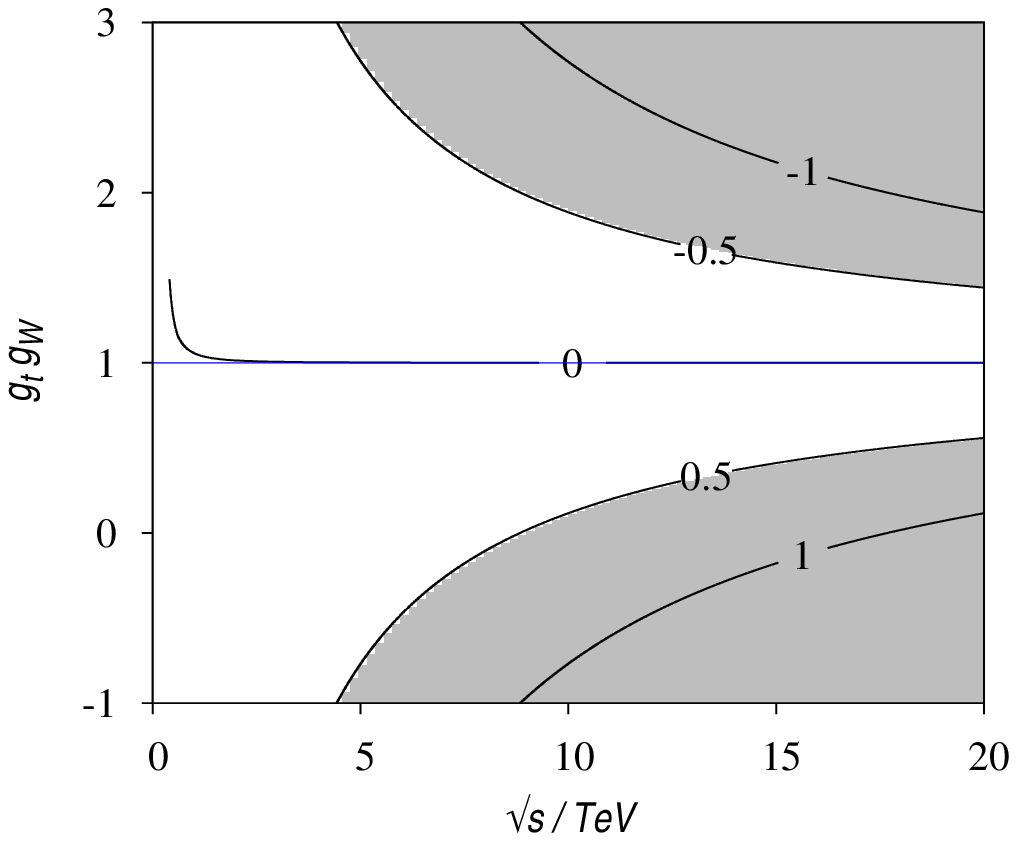}
\includegraphics[width=0.45\textwidth]{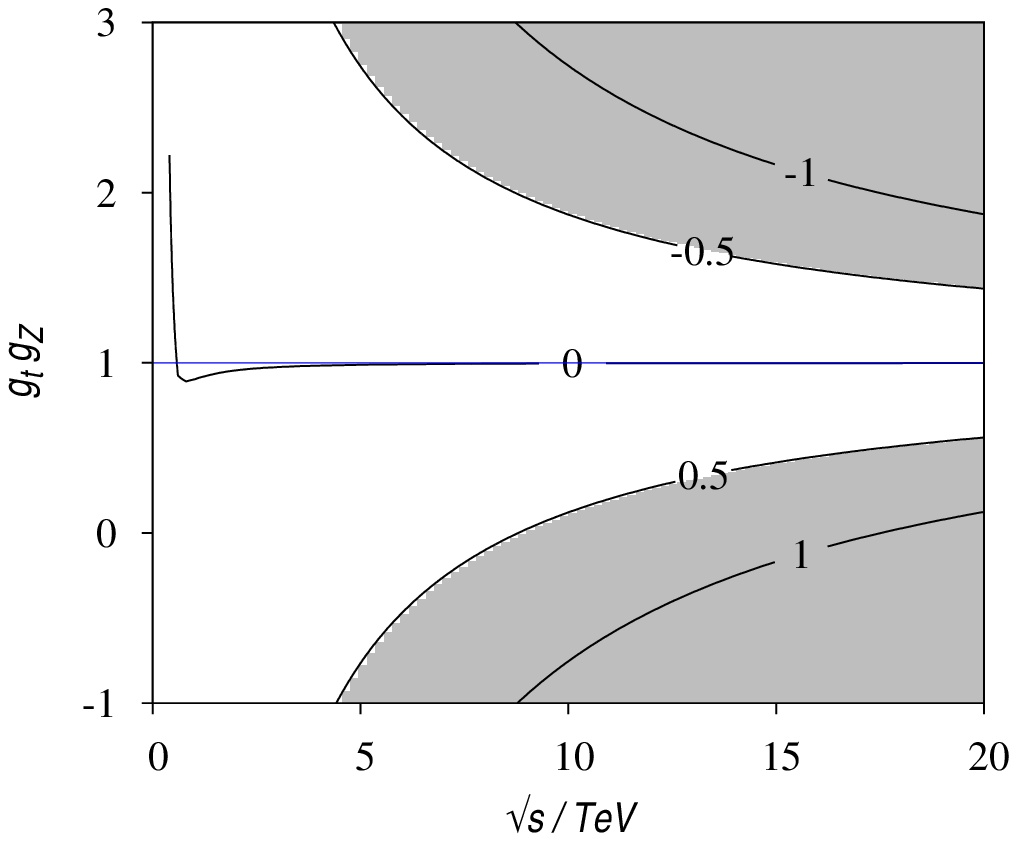}
\caption{\small \em Contours for $Re[a_0(0,0,1,1)]$
in the $g_t g_V$--$\sqrt{s}$ plane. The left (right) panel
corresponds to $W^+_L W^-_L \to t_+ \bar t_+$ ($Z_L Z_L \to t_+ \bar t_+$).
All couplings other than $g_t$ and $g_V$ are held
to the SM values.}
\label{fig:Contours}
\end{figure*}

In Fig.~\ref{fig:Contours} we display this data in terms
of iso-$Re(a_0)$ contours in the $g_t g_V$--$\sqrt{s}$ plane. Only the
white part of the figures bounded by the curves $Re(a_0) = \pm 0.5$ are in
consonance with unitarity, and the shaded regions are ruled out.
Once again, this shows that even if all the
other couplings were left unmolested, a large deviation in $g_t$ alone
would run afoul of unitarity constraints well within a few TeVs. This certainly
holds not only for the most favoured values quoted by
Ref.~\cite{Banerjee:2012xc} but  also for a very large fraction of their
$95\%$ C.L. allowed regions.

\section{Vacuum Stability}
\label{sec:rge}

If the consideration of unitarity persuades us to set some cut-off
scale to the SM augmented by the anomalous couplings, it is important
to check whether the theory respects all other constraints,
experimental as well as theoretical.  A case in point is the issue of
vacuum stability, which demands that the Higgs quartic coupling has to
be positive.  Once we allow the possibility of a deviation from the SM
Yukawa coupling, namely $g_t\not= 1$, the RG evolution for the Higgs
quartic coupling $\lambda$ would be affected too. 
The RG equation for
$\lambda$ involves only even powers of top Yukawa coupling $h_t$, which, 
with $\delta = 0$, is just $g_t$ times the SM top Yukawa coupling.
Thus the sign of $g_t$ is irrelevant, with the evolution depending only
on its magnitude.

We use the two-loop $\beta$-function for $\lambda$, following Refs.
\cite{RG-refs}. For completeness, they are also quoted in Appendix
\ref{app:betafn}. We use the two-loop matching conditions, as given in
Ref.~\cite{Bezrukov:2012sa},
to match the data, {\em viz.}
\beq
\barr{rclcrcl}
m_Z^{\rm pole} & = & 91.1876~{\rm GeV}\, &\qquad &
\alpha_s (m_Z) & = & 0.1184 \\
m_h^{\rm pole} & = & 125.3~{\rm GeV}\, & & 
\alpha (m_Z) &= &1 / 127.916
\\
m_t^{\rm pole} & = & 172.9~{\rm GeV} 
&& 
\s^2_W (m_Z) & = & 0.23116\,.
\earr
\eeq
to their corresponding values at $m_t$.

\begin{figure*}[!ht]
\centering
\includegraphics[width=0.45\textwidth]{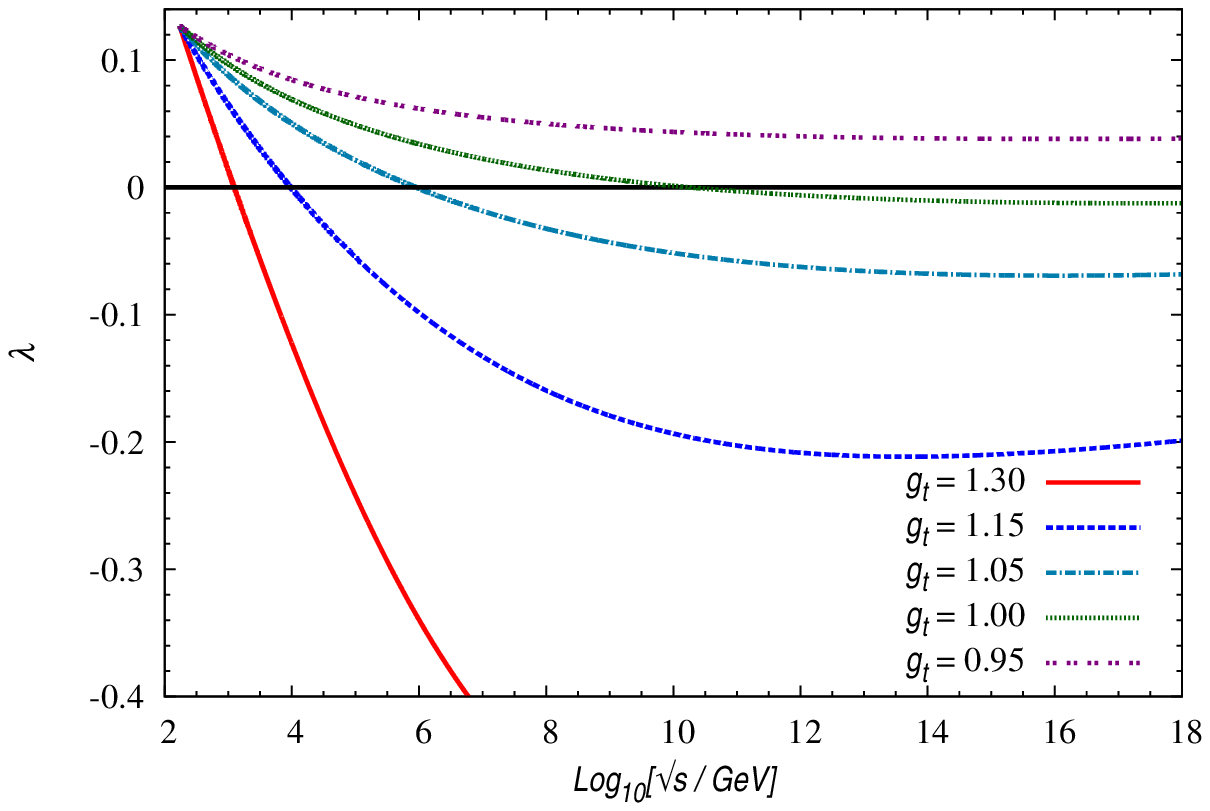}
\includegraphics[width=0.45\textwidth]{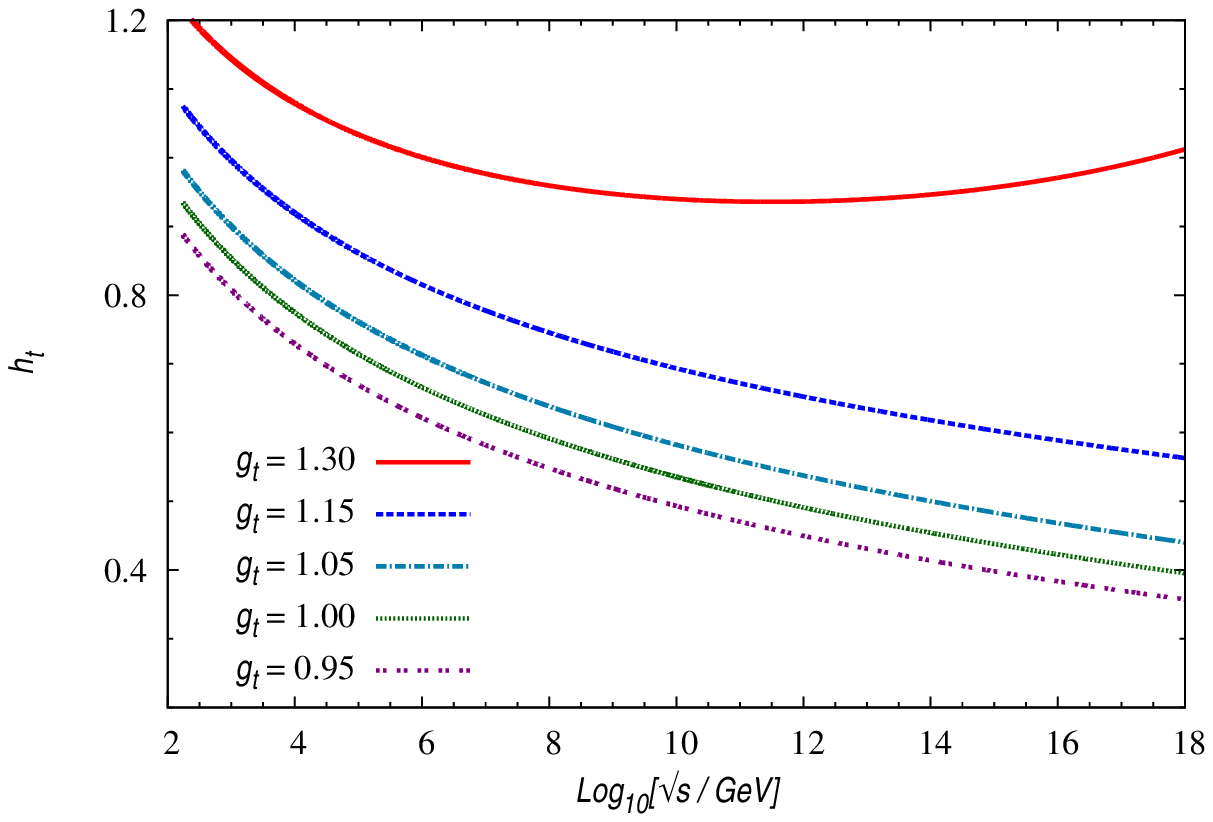}
\caption{\small \em {Variation of quartic Higgs coupling $\lambda$ and
the top quark Yukawa coupling $h_t$ with CM energy $\sqrt s$. Here $h_t$
is equal to $g_t$ times the SM top Yukawa coupling.}}

\label{fig:lambda}
\end{figure*}

We show the evolution of the scalar quartic coupling $\lambda$, and
the top Yukawa coupling, in Fig.~\ref{fig:lambda}. This shows that
the electroweak vacuum might get unstable if $|g_t|$ is even slightly
greater than unity, and the point where the instability sets in
depends rather sensitively on $|g_t|$. For example, the vacuum becomes
unstable at an energy as low as about $10^4$ GeV for $g_t = 1.15$. At
the one-loop level, the negative term proportional to $g_t^4$, coming
from a top-mediated box diagram, is responsible for this. Thus, both
F2 and F3 would indicate the presence of new physics $\sim 10^4$
GeV on this account (the unitarity bounds are stronger, though), while
F1 seems to be safe.  While these shifts parallel those engendered
by the errors on the top quark mass measurement itself, there are
subtle differences. For one, the shift in $|g_t|$ that the fittings
favour are much larger than the experimental errors in $m_t$
($0.6\%$--$1.5\%$ according to various estimates).  Moreover, the
deployment of the matching conditions in the two cases would differ.

At the same time, we must be cautious about taking these numbers too
literally.  The calculations hold only if the new physics responsible
for the change in the top Yukawa coupling is either above the scale
where instability sets in (so that those new degrees of freedom are
still frozen), or the effective interaction involves only SM fields
but with a new operator structure. In particular, the apparent consistency 
of F1 cannot be depended on, once the physics responsible for unitarity 
violation is turned on.

\section{Contemplating Possible Avenues}
\label{sec:avenue}

While we have delineated the problems that beset an effective theory
wherein the coupling of the recently glimpsed Higgs-like resonance to
the top quark, the $W$ and the $Z$ are reset from the SM values to
those obtained from phenomenological best fits, 
we have not indicated any source for the same. For example, the
  generation of a large (but real) anomalous coupling to be the result
  of either large quantum corrections or mixings with as yet
  undetected states. The latter possibility would, of course, require
  such states to be relatively low-lying, and in fact not too
  separated from the corresponding known SM states. Were it indeed to
  be so, all arguments about unitarity or triviality would necessarily
  need to be revised. Since
  an exhaustive treatment is not possible
  owing to the paucity of independent data, as well as the enormity of
  the task, we examine some simple alternatives. Viewed differently, 
  while the arguments in the previous sections point to the {\em necessity}
  of having a relatively low cutoff, we now consider some possible 
  realizations of the same.

Before we delve into the specifics, let us consider
some generic issues. 
\begin{itemize}
\item It might be argued that a negative $g_t$ would necessitate the
  existence of a second source of electroweak symmetry breaking and
  that this would imply relatively light new states. However, a close
  examination of the same shows that while the first part of the
  argument does hold, the second depends on implicit assumptions about
  the new sector. For example, one could well admit a second higgs
  doublet or a strongly interacting sector (such as topcolor), perhaps
  coupling only to the top quark. At the cost of some fine tuning (or 
  introduction of additional symmetries, analogous but not identical to 
  those in Little Higgs models), one
  could easily raise the mass scale for the new particles to a few
  TeVs. What the preceding arguments do show, however, that,
  independent of the fine tuning, this scale has to be lower than
  $\sim {\cal O}(10)$ TeV.
\item 
The introduction of any such ultraviolet completion would have an
impact on the running of the Higgs quartic couplings. This is
irrespective of whether this new sector couples to the SM Higgs at the
tree level. This would, presumably, cure the potential problem with
vacuum stability. However, unless the entire theory is known, an
accurate computation of this effect is not possible, and we will not
attempt to do so.
\end{itemize}

\subsection{Gauge boson scattering}

To begin with, let us consider the effect of $g_{W / Z} \neq 1$. While
unequal values for $g_W$ and $g_Z$ do violate custodial symmetry, the
most visible consequences appear in electroweak precision
observables and can be neutralized by arranging for compensating
custodial breaking in other sectors of the theory. Indeed, this 
has been included 
in the fitting of Ref.~\cite{Banerjee:2012xc}. As for the unitarity
violation in gauge boson scattering, curing it would require the
introduction of additional contributions to the amplitude. The
simplest possibility\footnote{As explained earlier, we do not consider
modification of the gauge boson self couplings.}  would be to
postulate the existence of another scalar, say $\tilde h$, whose
couplings parallel those of $h$ in Eq.~(\ref{biswa_effective}), but
with the corresponding couplings being $\tilde g_i$, viz.
\beq
     g_W \to  \tilde g_W\,, \qquad
     g_Z \to \tilde g_Z \, .
\eeq
Assuming that this new scalar has a mass $\tilde M \gg m_h$,
the restoration of unitarity for $\sqrt{s} \gg \tilde M$ would require
that
\beq
     \tilde g_W^2 + g_W^2 = 1 \ , \qquad
     \tilde g_Z^2 + g_Z^2 = 1 \ , \qquad
     \tilde g_W \, \tilde g_Z + g_W \, g_Z = 1 \ ,
 \label{restoration_1}
\eeq
with the three constraints emanating from considerations of $W_L^+
W_L^- \to W_L^+ W_L^-$, $Z_L Z_L \to Z_L Z_L$ and $W^+_L W^-_L \to Z_L
Z_L$ (and crossed processes) respectively.  While the requirements
might seem trivial at first sight, note that these are actually three
conditions on two variables. Moreover, the ``best fit points'' and,
indeed, most of the good fit part of the parameter space found in
\cite{Banerjee:2012xc} requires $|g_{W/Z}|^2 > 1$, thereby
necessitating negative $ | \tilde g_{W/Z}|^2$.  While this roadblock
could be circumvented by postulating a wrong sign for the scalar
kinetic term, such a solution brings along its own problems.
Note, though, that this would still not guarantee the existence of
a simultaneous solution to all three of the above
constraints. However, the extent of unitarity violation could be
minimized so as to push the scale of violation significantly higher.

The situation simplifies considerably if the scalar $\tilde h$ is not
an {\em ad hoc} degree of freedom, but part of another Higgs multiplet
that contributes to electroweak symmetry breaking. While only certain
representations would guarantee $m_W^2 = m_Z^2 \, \cos^2 \theta_W$ at
the tree-level, it is possible, in principle, to arrange multiple
vacuum expectation values for a multitude of representations and
carefully tune them to maintain this relation
\cite{Gunion:1989we,Choudhury:2003ut}. Obtaining effective $g_{W/Z} >
1$ for at least one such scalar (to be identified with the observed
resonance) requires that at least one of these representations must be
higher than a doublet \cite{Choudhury:2003ut}. Typically, though, $g_W
= g_Z$ would not be maintained. It should be realized that, now, it is
not just one new scalar that we would have, but an entire multiplet.
This, of course, would change eqs.(\ref{restoration_1}) to include
additional terms, thereby making it easier to satisfy all three
conditions. (This is despite the fact that gauge symmetry would relate
several of the new couplings.) The behaviour of the potentially
offending cross sections (equivalently, the partial wave amplitudes)
would change too; interim phases of growth with $\sqrt{s}$ would be
seen, especially as a new Higgs threshold is approached. For very
large $\sqrt{s}$ though, the Froissart bound would be seen to be
validated.

Yet another way to obtain $g_{W/Z} > 1$ is to postulate a new scalar
with non-standard kinetic terms for at least one of the two (the new scalar
and $h$) such that significant kinetic mixing occurs. An example of this
is afforded by the radion in warped models~\cite{Choudhury:2003ut}.

From scalars, we turn our attention to vector bosons as
restorers. Unitarizing gauge boson scattering in Higgs-less models
through the introduction of new vector bosons has been investigated in
Refs.~\cite{Csaki:2003dt,Nomura:2003du}.  Clearly, the couplings must
satisfy certain conditions. In the presence of a Higgs (albeit with
altered couplings), the relations of
Ref.~\cite{Csaki:2003dt,Nomura:2003du} have to be altered suitably.
The required changes are straightforward, at least as far as the
scattering of the SM gauge bosons is concerned. It must be noted,
though, that the introduction of such vector bosons introduces the
possibility of a pair of them emanating from, say, $W_L^+ W_L^-$
annihilation. The latter scattering would be associated with its own
unitarity violation problems, and, just as in the case of the
Higgs-less models, one would have to introduce a tower of such gauge
bosons. The tower, in principle, is an infinite one and can be
truncated only at the cost of admitting unitarity violation at some
scale (or, equivalently, appealing to some ultraviolet
completion). Similarly, all the trilinear (and quartic) couplings
between this set of vector bosons must satisfy sum rules, the
character of which will depend on whether they couple to $\tilde h$.

Finally, it should be noted that this role of unitarity restoration
is not restricted to only scalars and vector bosons, but can also
be assumed by higher-spin bosons. The inclusion of the latter, though,
brings a whole new set of problems to the table, and we desist from any
discussion of the same.

\subsection{Gauge boson annihilation to fermion pairs}

The introduction of a new scalar $\tilde h$ could, in principle, restore
unitarity for such processes. Once again,
denoting the coupling of $\tilde h$ to a $t \bar t$ pair through a
form analogous to Eq.~(\ref{biswa_effective}),
but with $g_t \to  \tilde g_t$, we can express
the conditions for unitarity restoration as
\beq
   \tilde g_t \, \tilde g_W + g_t \, g_W = 1 \ ,
\qquad
   \tilde g_t \, \tilde g_Z + g_t \, g_Z = 1 \ .
\eeq
As before, we are faced with the problem of simultaneous solution of
both these constraints, especially once $\tilde g_{W/Z}$ are
determined from considerations of gauge boson scattering (see
preceding section). Of course, with the amplitude here growing only
as $\sqrt{s} / m_W$, a lack of cancellation can be accommodated
to a relatively larger degree, yet postponing unitarity violation to
$\sqrt{s} > 10$ TeV, or even later.

The main problem, though, is that the best fit requires $g_t \,
g_{W/Z} \sim -1$. This, of course, entails having $\tilde g_t \,
\tilde g_{W/Z} \sim 2$, or, in other words, rather large couplings for
the $\tilde h$. Of particular importance is the fact that the
inclusion of scalars in larger representations of $SU(2)$ (and
ascribing vacuum expectation values to them) as in the preceding
section, not only does not help, but actually worsens the
situation. The reason is easy to see. As such large representations
would not couple to the top quark (barring non-renormalizable terms),
if the wavefunction of the observed resonance were to carry a
significant fraction of such a state, its coupling to the top would
actually be reduced from the SM value. This, of course, goes against
current observations.  Thus, one is left with the problem of arranging
a large $\tilde g_t$.  Within the ambit of a phenomenological
Lagrangian, this is admissible and can be arranged by invoking a
suitably large Yukawa coupling (renormalizable if the extra higgs
field is a doublet, and nonrenormalizable otherwise). However, note,
that such a Yukawa coupling would grow rapidly with energy and one
would be faced with a Landau pole.  An alternative could be to
consider new fermions or gauge bosons which may not couple to our
familiar Higgs doublet.  However, it is easy to see that this does not
help as long as the latter behave canonically.

To summarise, the introduction of a new (set of) scalars with
carefully constructed couplings seems to offer the simplest solution
to the conundrum. A strictly phenomenological approach, on the other
hand, would be given by ascribing form factor behaviour to the
deviations.  For example, consider the replacement
\[
  \delta g_i \to \delta g_i^0 \, 
\left(\dfrac{2 \, m_h^2}{s + m_h^2}\right)^{n_i} \ , \qquad n_i \geq 1 \ , \qquad i = t, W, Z
\]
This, clearly would restore unitarity at large energies. This has the
further advantage that this permits an examination of the behaviour of
the coupling at different energies, thereby permitting some insight
into the structure of the deviation once more data is available. Of
course, a more generic form factor can be used instead, even
correcting for the lack of gauge invariance that the simple-minded
expression above entails. This, though, takes us to the regime of
electroweak chiral Lagrangians and we shall not delve into it any
further.

\section{Future Outlook and Conclusions}
\label{sec:conc}

Assuming that the Higgs couplings to the SM fields are arbitrary but
consistent with general principles like Lorentz invariance and
hermiticity, we tried to see whether the present data gives any hint
of new physics beyond the SM.  A particularly sensitive probe is
offered by considerations of unitarity in gauge boson scattering.  We
have considered several such scattering amplitudes, for polarized as
well as unpolarized gauge bosons, and partial wave unitarity is seen
to break down at about $\sqrt{s} \, \gtap 4$ TeV for coupling values
preferred by the fits. 

Even if this can be prevented by restoring the $hWW$ and $hZZ$
vertices to their SM values (especially since the best fits, anyway,
call for only small deviations), we are still faced two rather
interesting issues. Indeed, the most important parameter in the
study is the top quark Yukawa coupling, 
which might even have
a sign opposite to that of the SM prediction.
because of the apparent excess of Higgs to diphoton decay rate.
 We explored the consequences of such a wrong-sign coupling.

There are two places where the wrong-sign Yukawa coupling can play
havoc. The first is the unitarity in gauge boson annihilation to 
a $t \bar t$ pair. 
 This effect can be traced to a term in the
scattering amplitude which is proportional to the product of the top
quark Yukawa coupling and the $hVV$ coupling.  With the sign flip of
this term, the amplitudes grow up instead of going down and one sees 
unitarity violation at  $\sqrt{s} \, \gtap 5$ TeV. Thus, this
indicates some new physics which takes over at a few TeV scale and
restores unitarity as well as gauge invariance, which is apparently
broken by Eq.~(\ref{our_effective}).

The second place is the stability of the electroweak vacuum. The Higgs
quartic coupling $\lambda$ becomes negative if the magnitude of the 
top Yukawa coupling
increases even a little from its SM value (only $\vert
g_t\vert$ is important here, and not the sign of $g_t$). The point where the vacuum
becomes unstable is a sensitive function of $g_t$, but for our
benchmark points, occur between 1 and 10 TeV, a region already
indicated by the unitarity violation. Again, this asks for some new
degrees of freedom, which couple to the Higgs and make the vacuum
stable (so these should better be bosonic in nature).  Of course,
whether the cutoff of the theory is at the Planck scale or at a few
TeVs does not affect the $h\to\gamma\gamma$ rate as this must always be
finite.

It might be argued that the departures from standard couplings as
suggested by the data are based on {\em global analyses}, where other
couplings are simultaneously assuming non-standard values. This could
be construed to mean that a complete analysis will have to take into
account the role of the other modified couplings in the evolution of
$\lambda$ as well as in ensuring unitarity in scattering
phenomena. While, as a principle, this is certainly true, note that
our analysis has included all of the relevant dimension-4 terms that
can be written down in terms of the SM fields alone. Although the
inclusion of subdominant terms would alter the quantitative details of
our conclusions, no qualitative change would be brought about.

Thus, if the initial trend---in particular the excess in diphoton
channel---persists in the new data, this might lead to some indirect
evidence of new physics which is lurking close.  It is worthwhile to
consider the possibility that the accumulation of further luminosity
would reduce the discrepancy between the data and the SM expectations,
without completely obliterating it.  Were this to be the case, one
would still need small but non-zero values for one or more anomalous
couplings. A direct measurement of such small changes in the effective
couplings would be well-nigh impossible, especially in the LHC
environment. Pending future experiments, considerations of unitarity
and vacuum instability would then consist of the best ``evidence'' for
a relatively low-lying threshold. However, if the effective couplings
shift by less than 10\% of their SM values, the minimum required scale
for new physics would rise to $\sim 100$ TeV. 

Recently, both ATLAS and CMS have updated their results for Higgs
search at $\sqrt{s} = 8$ TeV. These include not only those for the
channels that form the core of our analysis, namely $h \to
\gamma\gamma$ ~\cite{ATLASnew12,CMSnew1}, $h \to ZZ^*$
~\cite{ATLAS_coup, CMSnew2} and $h \to WW^*$ ~\cite{ATLASnew30,
  CMSnew5}, but also for others such as $h \to Z
\gamma$~\cite{ATLASnew9, CMSnew6}. Based on these, it has been
variously claimed that the observations are almost perfectly in
consonance with the SM expectations. However, a careful examination
betrays a persisting lack of consistency between the various
measurements. The large variations in the data, alongwith the data in
$h \to b \bar b$~\cite{ATLASold170, CMSold44} as well as $h \to \tau^+
\tau^-$~\cite{ATLASold160,CMSnew4} has led to subsequent
fits~\cite{falkowski, giardino}.  Although these are in the spirit of
earlier fits~\cite{fits-past, fits-recent, Banerjee:2012xc}, the
incorporation of new data has led to a shift in the best fit values
for the couplings somewhat away from those of
Ref.~\cite{Banerjee:2012xc}. Notwithstanding these changes, the most
important message is that the values for $g_{t,W,Z}$ can yet be far
away from those within the SM. In particular, $g_t$ can be
substantially different from unity~\cite{falkowski, giardino} (with
the central issue of this paper still remaining a concern).  It should
be appreciated, though, that these are yet early days of Higgs physics
and the central values may yet shift!

\vskip 10pt
\noindent
{\em Note added.}

While the work was being completed, we became aware of a similar work
 in progress \cite{dipankar}.


\vskip 5mm

\acknowledgments

The authors thank Biswarup Mukhopadhyaya for collaboration 
during the early phase of the investigation and 
G. Rajasekaran for an insightful comment.
D.C. thanks Debashis Ghoshal for useful discussions.
R.I. was supported by CSIR, Government of India, 
under grant 09/045(0872)/2009-EMR-I.
A.K. was supported by CSIR, Government of India (project no.\
03(1135)/9/EMR-II), and also by the DRS programme of the UGC,
Government of India.  


\appendix
\section{Momenta, Polarizations and Helicity Amplitudes}
In our calculations we have denoted the momenta of the particles as follows
\bea\label{mom}
k_1 = \dfrac {\sqrt{s}}{2} (1,0,0,\beta_V);
& &
k_2 = \dfrac {\sqrt{s}}{2} (1,0,0,-\beta_V);
\notag\\
k_3 = \dfrac {\sqrt{s}}{2} (1,\beta_t \s_\theta,0,\beta_t \c_\theta);
& &
k_4 = \dfrac {\sqrt{s}}{2} (1,- \beta_t \s_\theta,0,- \beta_t \c_\theta),
\eea
where $\sqrt s$ is the CM energy, $\beta_V = \sqrt{1 - 4m^2_V/s}$ and
$\beta_t = \sqrt{1 - 4m^2_t/s}$. $V=W^\pm,Z$ in the appropriate cases.

The polarization vectors have been denoted as
\bea\label{pol-V}
\e^{\hat \lambda}_{k_1} = \dfrac{1}{\sqrt2} (-{\hat \lambda}\e_1 - i{\hat \lambda}^2\e_2) + (1-{\hat \lambda}^2) \e_3;
\qquad
\e^{\hat \lambda}_{k_2} = \dfrac{1}{\sqrt2} ( {\hat \lambda}\e_1 - i{\hat \lambda}^2\e_2) + (1-{\hat \lambda}^2) \e_4;
\eea
where ${\hat \lambda} = 0$ corresponds to the longitudinal and ${\hat \lambda}=\pm$
are the transverse polarizations. $\e_i$ are as follows
\bea
\e_1 = (0,1,0,0);\ \
\e_2 = (0,0,1,0);\ \
\e_3 = \dfrac{\sqrt s}{2 m_V} (\beta_V,0,0,1);\ \ 
\e_4 = \dfrac{\sqrt s}{2 m_V} (\beta_V,0,0,-1).
\eea

The helicity states of top quarks are given by
\bea\label{hel-top}
\chi_+(k_3) =
  \begin{pmatrix}
  \c_{\theta/2} \\ \s_{\theta/2}
  \end{pmatrix},
   \ \
\chi_-(k_3) =
  \begin{pmatrix}
  - \s_{\theta/2} \\ \c_{\theta/2}
  \end{pmatrix} ;
\ \
\chi_+(k_4) =
  \begin{pmatrix}
  \s_{\theta/2} \\ -\c_{\theta/2}
  \end{pmatrix},
   \ \
\chi_-(k_4) =
  \begin{pmatrix}
  \c_{\theta/2} \\ \s_{\theta/2}
  \end{pmatrix}.
\eea
From \eqref{hel-top} we can get the 4-component Dirac spinors as
\beq
u (p,{\hat \lambda}) =
\begin{pmatrix} \omega_{-{\hat \lambda}}(p)\ \chi_{\hat \lambda} 
(\hat {\bf p}) \\ \omega_{\hat \lambda}(p)\ \chi_{\hat \lambda} (\hat {\bf p}) \end{pmatrix};
\qquad
v (p,{\hat \lambda}) =
\begin{pmatrix} -{\hat \lambda}\omega_{\hat \lambda}(p)\ \chi_{-{\hat \lambda}} 
(\hat {\bf p}) \\ {\hat \lambda} \omega_{-{\hat \lambda}}(p)\ \chi_{-{\hat \lambda}} (\hat {\bf p}) \end{pmatrix}.
\eeq
Here we have defined $\omega_{\hat \lambda}(p) = \sqrt{E + {\hat \lambda} |{\bf p}|}$.

Using the momenta from \eqref{mom}, polarizations and helicity states
from \eqref{pol-V}, \eqref{hel-top} respectively and the taking the
effective Lagrangian of Eq.~(\ref{biswa_effective}), we get the helicity
amplitudes for $W^+W^- \to t\bar t$ as
\begin{eqnarray}
 \label{WWgs}
{\cal M}^{\gamma s}_{0011} &=&
- \dfrac23 g^2 \s^2_W \dfrac{1}{s} \dfrac{m_t}{m_W^2} \beta_W \sqrt s \left(s+2 m_W^2\right) \c_{\theta}\,,
\nonumber\\
{\cal M}^{Z s}_{0011} &=&
- g^2 \left[1 - \dfrac83\s^2_W\right] \dfrac{1}{s - m^2_W} \dfrac{m_t}{4 m_W^2} \beta_W \sqrt s \left(s+2 m_W^2\right) \c_{\theta}\,,
\nonumber\\
{\cal M}^t_{0011} &=&
- \dfrac{g^2}{2} |V_{tb}|^2 \dfrac{1}{t} \dfrac{m_t}{8m_W^2} s^{3/2} [\beta_W(1-\beta^2_W)\c_\theta - 2\beta_t \c^2_\theta + 
\beta_t (1+\beta^2_W)]\,,\nonumber\\
{\cal M}^h_{0011} &=&
- g_t g_W \dfrac{g^2 m_t}{2} \dfrac{1}{s - m_h^2} \dfrac12\beta_t \sqrt s \left( \dfrac{s}{m^2_W} - 2 \right)
\end{eqnarray}

Similarly we get the helicity amplitudes for $ZZ \to t\bar t$ as
\begin{eqnarray} 
 \label{ZZt}
{\cal M}^t_{0011} &=&
-[(g^{Zt}_L)^2 + (g^{Zt}_R)^2] \dfrac{1}{t-m^2_t} \dfrac{m_t}{8m_Z^2} s^{3/2} 
[\beta_Z(1-\beta^2_Z)\c_\theta - 2\beta_t \c^2_\theta + \beta_t (1+\beta^2_Z)] \nonumber\\
&& +
g^{Zt}_L g^{Zt}_R \dfrac{1}{t-m^2_t} \dfrac{m_t}{4m^2_Z} s^{3/2} [\beta_t(1+\beta^2_Z) - 2\beta_Z\c_\theta]\,,
\nonumber\\
{\cal M}^u_{0011} &=&
-[(g^{Zt}_L)^2 + (g^{Zt}_R)^2] \dfrac{1}{u-m^2_t} \dfrac{m_t}{8m_Z^2} s^{3/2} 
[- \beta_Z(1-\beta^2_Z)\c_\theta - 2\beta_t \c^2_\theta + \beta_t (1+\beta^2_Z)]
\nonumber\\
&& +
g^{Zt}_L g^{Zt}_R \dfrac{1}{u-m^2_t} \dfrac{m_t}{4m^2_Z} s^{3/2} [\beta_t(1+\beta^2_Z) + 2\beta_Z\c_\theta]\,,
\nonumber\\
{\cal M}^h_{0011} &=&
- g_t g_Z \dfrac{g^2 m_t}{4 \c^2_W} \dfrac{1}{s - m_h^2} \beta_t \sqrt s \left( \dfrac{s}{m^2_Z} - 2 \right)\,,
\end{eqnarray}
where $ g^{Zt}_L = - \dfrac{g}{2\c_W} \left(1 - \dfrac43\s^2_W\right), \
g^{Zt}_R = \dfrac{g}{2\c_W} \dfrac43\s^2_W $. $\s_W^2 = \sin^2\theta_W$,
$\c_W^2 = \cos^2\theta_W$ and $\theta_W$, the Weinberg angle.

\section{Beta Functions}\label{app:betafn}
We give the beta functions used in our calculation from the appendix of
Ref.~\cite{Holthausen:2011aa}. The beta function for a generic coupling 
\(X\) is given as:
\beq
\mu\dfrac{\mathrm{d}X}{\mathrm{d}\mu} = 
\beta_X =\sum_i\dfrac{\beta_X^{(i)}}{(16\pi^2)^i}\,.
\eeq
The beta functions are given, above $m_t$ but below any new degrees of freedom,
by \cite{RG-refs}:
\bea
\beta_{\lambda}^{(1)} &=&
  24 \, \lambda^2
  + 12 \, \lambda \, h_t^2
  - 6 \, h_t^4
  - 3 \, \lambda \, g^2_1
  - 9 \, \lambda \, g^2_2
  + \dfrac{3}{4} \, g_2^4
  + \dfrac{3}{8} \, (g_1^2+g_2^2)^2\,,
\nonumber\\
\beta_{h_t}^{(1)} &=&
  \dfrac{9}{2} \, h_t^3
  - \dfrac{17}{12} \, h_t \, g_1^2
  - \dfrac{9}{4} \, h_t \, g_2^2
  - 8 \, h_t \, g_3^2\,,
\nonumber\\
\beta_{g_1}^{(1)} &=&
  \dfrac{41}{6} \, g_1^3,
\qquad
\beta_{g_2}^{(1)} =
  - \dfrac{19}{6} \, g_2^3,
\qquad
\beta_{g_3}^{(1)} =
  - 7 \, g_3^3\,,
\nonumber\\
\beta_{\lambda}^{(2)} &=&
  - 312 \, \lambda^3
  - 144 \, \lambda^2 h_t^2
  - 3 \, \lambda h_t^4
  + 36 \, \lambda^2 \, g_1^2
  + 108 \, \lambda^2 \, g_2^2
  + 80 \, \lambda h_t^2 \, g_3^2
  + \dfrac{45}{2} \, \lambda h_t^2 \, g_2^2
  + \dfrac{85}{6} \, \lambda h_t^2 \, g_1^2
\nonumber\\
  & &
  - \dfrac{73}{8} \, \lambda g_2^4
  + \dfrac{39}{4} \, \lambda g_2^2 \, g_1^2
  + \dfrac{629}{24} \, \lambda g_1^4
  + 30 \, h_t^6
  - 32 \, h_t^4 \, g_3^2
  - \dfrac{8}{3} \, h_t^4 \, g_1^2
  - \dfrac{9}{4} \, h_t^2 \, g_2^4
\nonumber \\
  & &
  + \dfrac{21}{2} \, h_t^2 \, g_2^2 \, g_1^2
  - \dfrac{19}{4} \, h_t^2 \, g_1^4
  + \dfrac{305}{16} \, g_2^6
  - \dfrac{289}{48} \, g_2^4 \, g_1^2
  - \dfrac{559}{48} \, g_2^2 \, g_1^4
  - \dfrac{379}{48} \, g_1^6\,,
\nonumber\\
\beta_{h_t}^{(2)} &=&
  6 \, \lambda^2 \, h_t
  - 12 \, \lambda \, h_t^3
  - 12 \, h_t^5
  + \dfrac{131}{16} \, h_t^3 \, g_1^2
  + \dfrac{1187}{216} \, h_t \, g_1^4
  - \dfrac{3}{4} \, h_t \, g_1^2 \, g_2^2
  + \dfrac{19}{9} \, h_t \, g_1^2 \, g_3^2
  + \dfrac{225}{16} \, h_t^3 \, g_2^2
\nonumber\\
  & &
  - \dfrac{23}{4} \, h_t \, g_2^4
  + 9 \, h_t \, g_2^2 \, g_3^2
  + 36 \, h_t^3 \, g_3^2
  - 108 \, h_t \, g_3^4\,,
\nonumber\\
\beta_{g_1}^{(2)} &=&
  - \dfrac{17}{6} \, g_1^3 \, h_t^2
  + \dfrac{199}{18} \, g_1^5
  + \dfrac{9}{2} \, g_1^3 \, g_2^2
  + \dfrac{44}{3} \, g_1^3 \, g_3^2\,,
\nonumber\\
\beta_{g_2}^{(2)} &=&
  - \dfrac{3}{2} \, h_t^2 \, g_2^3
  + \dfrac{3}{2} \, g_1^2 \, g_2^3
  + \dfrac{35}{6} \, g_2^5
  + 12 \, g_2^3 \, g_3^2\,,
\nonumber\\
\beta_{g_3}^{(2)} &=&
  - 2 \, h_t^2 \, g_3^3
  + \dfrac{11}{6} \, g_1^2 \, g_3^3
  + \dfrac{9}{2} \, g_2^2 \, g_3^3
  - 26 \, g_3^5\,.
   \label{the_beta_fns}
\eea
In the above, $\lambda$ is the quartic Higgs coupling, $h_t$, the top
Yukawa coupling, $g_1, g_2$ and $g_3$ are the $U(1)_Y, SU(2)_L$ and $SU(3)_C$
couplings respectively. 
As the evolution of $h_t$ involves $h_t$ itself rather than the SM 
top Yukawa coupling $\sqrt{2} m_t / v = h_t / g_t$, no new RGE appears; 
the existence of a nontrivial $g_t$ manifests itself only through the
replacement $h_t^{SM} \to h_t = g_t \, h_t^{SM}$ in each of the
eqs.(\ref{the_beta_fns}).  Also note that our description is purely
phenomenological and no dynamic origin is ascribed to $g_t$ (doing so
would need a specific ultraviolet completion).


\end{document}